\documentclass[10pt,final,journal]{IEEEtran}
\usepackage{hyperref}
\usepackage{cite}
\usepackage{mathtools}
\usepackage{bm}
\usepackage{amssymb}
\usepackage{color}
\usepackage{diffcoeff}
\diffdef{pvrule}{op-symbol = \partial}
\DeclareMathOperator{\diag}{diag}
\def\BibTeX{{\rm B\kern-.05em{\sc i\kern-.025em b}\kern-.08em
    T\kern-.1667em\lower.7ex\hbox{E}\kern-.125emX}}
\begin{document}
\title{Long Wavelength Coherency in Well Connected Electric Power Networks}
\author{Julian Fritzsch,~\IEEEmembership{Graduate Student Member, IEEE}, and Philippe Jacquod,~\IEEEmembership{Member, IEEE}
    \thanks{This work has been supported by the Swiss National Science Foundation under grant 200020\_182050}%
    \thanks{J. Fritzsch, and Ph. Jacquod are with the Department of Quantum Matter Physics, University of Geneva, CH-1211 Geneva, Switzerland and
        the School of Engineering, University of Applied Sciences of Western Switzerland HES-SO, CH-1951 Sion, Switzerland.}%
    \thanks{Emails:\tt{\small{julian.fritzsch@etu.unige.ch, philippe.jacquod@unige.ch}}}}%
\maketitle
\begin{abstract}
    We investigate coherent oscillations in large scale transmission power grids,
    where large groups of generators respond in unison to a distant disturbance. Such long wavelength coherent
    phenomena are known as inter-area oscillations. Their existence in networks of
    weakly connected areas is well captured by singular perturbation theory. However, they are also observed in strongly connected networks without time-scale separation, where applying singular perturbation theory is not justified. We show that the occurrence of these oscillations is actually generic. Applying matrix perturbation theory, we show that, because
    these modes have the lowest oscillation frequencies of the system,
    they are only moderately sensitive to increased network connectivity between
    well chosen, initially weakly connected areas, and that their general structure remains the same, regardless of the strength of the inter-area coupling.
    This is qualitatively understood by
    bringing together the standard singular perturbation theory and Courant's nodal
    domain theorem.
\end{abstract}
\section{Introduction}
Synchronous generators in interconnected AC electric power systems exhibit electro-mechanical oscillations. Of particular
interest are large-scale cooperative phenomena termed {\it inter-area oscillations} which are coherent,
sub-{\it Hz} frequency
oscillations between geographically separated large groups of generators~\cite{rogersPowerSystemOscillations2000}. Such oscillations
may become unstable
and lead to large scale blackouts~\cite{venkatasubramanianAnalysis1996Western2004}, therefore
safe system operation
requires that they are appropriately damped, which becomes harder as the energy transition unfolds.
As a matter of fact, power system stabilizers
installed on conventional synchronous generators have so far been the main source of damping against
inter-area oscillations~\cite{rogersPowerSystemOscillations2000} and
substituting new renewable sources of energy for conventional synchronous machines
reduces the availability of these resources. There is a vast literature on damping of inter-area oscillations
in power systems with large penetrations of new renewable generation, see e.g.~\cite{Zha16}.

\begin{figure}
    \centering
    \includegraphics{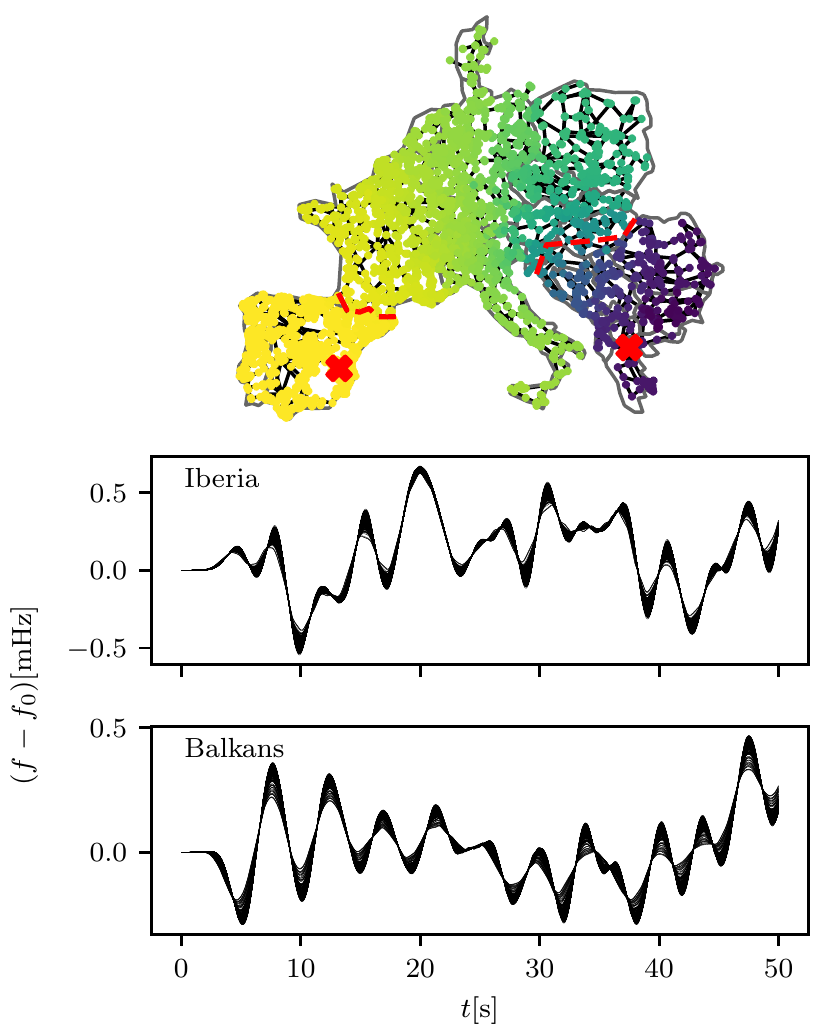}
    \caption{Top panel: linear combination of the slowest two modes of the Laplacian of the PanTaGruEl model~\cite{tylooKeyPlayerProblem2019,pagnierInertiaLocationSlow2019}. The nodal mode amplitudes are color-coded with maximal negative values in dark violet and maximal positive values in yellow. The dashed lines indicate the boundary of the Balkan and Iberian Peninsula areas that are connected via this inter-area mode.
        Bottom panels:
        Coherent responses of the 981 nodes in the Iberian Peninsula and the 368 nodes in the Balkans to a noisy power generation disturbance in the opposite area. The applied disturbance is an Ornstein-Uhlenbeck noise with a correlation time $\tau = 25\,\mathrm{s}$ and the disturbance locations are indicated by crosses in the top panel. The mode represented in the top panel is responsible for these coherent inter-area oscillations.}
    \label{fig:noise}
\end{figure}

To achieve optimal damping of inter-area oscillations it is important to first  identify the geographical areas that carry
them, i.e., where synchronous generators display the same frequency response following a fault
or other excitations. This identification commonly proceeds through highlighting
weak links~\cite{nathCoherencyBasedSystem1985}, spectral analysis~\cite{chowTimeScaleModelingDynamic1982}, and
data-based approaches identifying generators with similar frequency responses, either in simulations of the linearized dynamics~\cite{podmoreIdentificationCoherentGenerators1978} or using wide area measurement data~\cite{jonssonSystemProtectionScheme2004}. Once coherent areas are identified, inter-area oscillations
are studied using aggregated models constructed from singular perturbation theory~\cite{dateAggregationPropertiesLinearized1991,chowSingularPerturbationAnalysis1978,chowTimeScaleModelingDynamic1982}.
These methods presuppose the existence of at least one small parameter $\mu$ measuring the
ratio between the inter-area and the intra-area connection strength or the associated time scales.
When $\mu$ is not small, the theory loses its validity, yet
inter-area oscillations are observed even in networks with large $\mu$.
This is illustrated in Fig.~\ref{fig:noise} which shows coherent, low-frequency inter-area
oscillations obtained numerically in the PanTaGruEl model of the synchronous grid of continental Europe~\cite{tylooKeyPlayerProblem2019,pagnierInertiaLocationSlow2019}. It is seen that
the Iberian Peninsula responds coherently to a noisy power injection in Greece, and reciprocally, even though $\mu$ is close to one hundred.
All 981 nodes in the Iberian Peninsula respond coherently -- with the same frequency and phase -- to a single-node, noisy perturbation in the Balkans, as do all 368 Balkan nodes to
a similar perturbation in the Iberian Peninsula.
The system of Fig.~\ref{fig:noise} operates well outside the regime of validity of the standard singular perturbation theory~\cite{dateAggregationPropertiesLinearized1991,chowSingularPerturbationAnalysis1978,chowTimeScaleModelingDynamic1982}.

Another viewpoint predicts the existence of inter-area oscillations, regardless of the network connectivity.
The dynamics of voltage angles in power systems is commonly modeled by the
swing equations, which are a set of coupled ordinary differential equations~\cite{machowskiPowerSystemDynamics2020}.
The coupling is determined by a graph
Laplacian matrix, whose eigenvectors naturally define coherent areas. As a matter of fact, Courant's nodal domain theorem
states that the $k^\mathrm{th}$ eigenfunction of an elliptic operator acting on a bounded domain $\Omega$ defines no more than
$k$ nodal subdomains of $\Omega$, where the eigenfunction does not change sign~\cite{Cou23}.
To make a long story short, the eigenmodes with lowest eigenvalues of an elliptic operator such as, e.g., a Laplacian operator,
define few large areas -- nodal subdomains -- on which the sign of their components does not change.
When a single or very few eigenmodes are excited, the resulting
oscillations appear coherent inside the corresponding nodal domains.
The theorem has recently been extended from continuous elliptic operators to graphs represented by, e.g., a discrete
Laplacian matrix~\cite{Urs18}. In the case of power systems, that Laplacian matrix represents a quasi-planar graph and
nodal domains are two-dimensional areas. Courant's nodal domain theorem states that these areas are larger for slower
eigenvectors of the graph Laplacian -- those with lower frequencies.
Consequently, these modes do not resolve inhomogeneities
in the inertia and damping parameters, therefore the structure  of the system's true inter-area modes is directly determined by the
slowest eigenvectors of the Laplacian.
Low frequency coherent oscillations over large areas thus
naturally emerge from a modal decomposition of the graph Laplacian.

Our purpose in this manuscript is to connect the standard singular perturbation theory approach to inter-area oscillations
to this modal point of view, valid regardless of inter-area coupling. We will use matrix perturbation theory~\cite{Bam20} to describe low frequency oscillations, extending our earlier work~\cite{fritzschMatrixPerturbationTheory2021}. We will show how an appropriate
choice of
areas gives an excellent approximation for the modes responsible for inter-area oscillations, which is only very weakly sensitive to the
inter-area coupling. This fills an important gap in the theory of inter-area oscillations as it explains their persistence in strongly connected
power networks.

The manuscript is organized as follows.
Section~\ref{sec:model} introduces the network model and motivates the focus on the network Laplacian to describe the inter-area oscillations.
Matrix perturbation theory is discussed in Section~\ref{sec:pt}.
After a summary of the general results of perturbation theory we define our model for describing inter-area oscillations.
In the rest of the section the validity of perturbation theory is discussed in terms of series convergence and the change of eigenvectors due to avoided crossings.
Applications of the theory to a synthetic two-area network, the IEEE RTS 96 Test System, and the PanTaGruEl model of the synchronous very high voltage grid of continental Europe are shown in Section~\ref{sec:results}.
Results are discussed in Section~\ref{sec:conclusions}.

\section{Dynamical Model}\label{sec:model}
\subsection{Swing Equations}
We use the structure-preserving model of Ref.~\cite{bergenStructurePreservingModel1981} and consider the voltage angle dynamics of
a high voltage power grid with $N$ nodes. The dynamics of the voltage angle $\theta_i$ on generator nodes is determined by the swing equations~\cite{machowskiPowerSystemDynamics2020}. In the case of high voltage transmission grids, a standard approximation is
the lossless line approximation, which neglects Ohmic losses. The swing equations then read
\begin{equation}\label{eq:generators}
    m_i \ddot{\theta}_i + d_i \dot{\theta}_i = P_i - \sum_{j}B_{ij}\sin(\theta_i-\theta_j) \, ,
\end{equation}
with the inertia $m_i$ and damping $d_i$ parameters of the generator and their active power $P_i>0$. Loads are assumed
frequency-dependent, and the power they draw is given by a constant $P_i < 0$ and a frequency-dependent term, $d_i\dot{\theta}_i$.
Load voltage angles then obey~\cite{bergenStructurePreservingModel1981}
\begin{equation}\label{eq:loads}
    d_i\dot{\theta}_i = P_i- \sum_{j}B_{ij} \sin(\theta_i - \theta_j) \, ,
\end{equation}
where the frequency dependence of the loads is determined by the parameter~\cite{bergenStructurePreservingModel1981}
\begin{equation}
    d_i = \frac{\alpha}{\omega_0}\lvert P_i^{0} \rvert,
\end{equation}
with $P_i^{0}$ the power consumption at the nominal frequency $\omega_0$.
The parameter $\alpha$ has been evaluated experimentally, $\alpha \in [0.8,2]$~\cite{Welfonder_1989,O_Sullivan_1996}.
In Eqs.~\eqref{eq:generators} and \eqref{eq:loads}, $B_{ij}$ denotes the product of the voltage magnitudes at nodes $i$ and $j$ with
the line susceptance. In the lossless line approximation, line conductances are neglected.

We investigate the generation of inter-area oscillations through small signal stability analysis. Accordingly, we linearize
Eqs.~\eqref{eq:generators} and \eqref{eq:loads} about the operational synchronous state with
$\theta_i = \theta^{(0)}_i + \delta\theta_i$ and $P_i = P_i^{(0)} + \delta P_i$,
\begin{equation}\label{eq:swinglin}
    {\bm M} \delta \ddot{\bm{\theta}} +   {\bm D} \delta \dot{\bm{\theta}} = \delta {\bm P} - {\bm L} \delta \bm{\theta} \, ,
\end{equation}
where we grouped the voltage angle deviations into a vector $\delta \bm \theta$, and introduced the diagonal inertia and damping matrices,
${\bm M} = \diag(m_i)$ (with $m_i=0$ on load nodes), ${\bm D} = \diag(d_i)$ as well as the network Laplacian matrix ${\bm L}$,
\begin{equation}
    {\bm L}_{ij} = \begin{cases}
        -B_{ij} \cos(\theta_i^{(0)} - \theta_j^{(0)})     & \text{for } i \neq j \\
        \sum_{k}B_{ik}\cos(\theta_i^{(0)}-\theta_j^{(0)}) & \text{for } i = j
    \end{cases}.
\end{equation}
Eq.~\eqref{eq:swinglin} is often written as a linear system of ordinary differential equations with stability matrix ${\bm A}$,
\begin{equation}\label{eq:Ax}
    \dot{\bm{x}} = \bm{A}\bm{x} + \bm{\Pi},
\end{equation}
where angles, frequencies and power injections are grouped into $\bm x$ and $\bm{\Pi}$ as
\begin{equation}
    \bm{x} =\begin{bmatrix}
        \delta\bm{\theta}_g \\\delta\bm{\theta}_l\\\delta\dot{\bm{\theta}}_g
    \end{bmatrix}, \qquad \bm{\Pi} = \begin{bmatrix}
        \bm{0} \\\delta\bm{P}_l\\\delta\bm{P}_g
    \end{bmatrix},
\end{equation}
which defines the $\bm{A}$-matrix
\begin{equation}
    \bm{A} = \begin{bmatrix}
        \bm{0}                         & \bm{0}                          & \mathbb{I}                     \\
        -{\bm D}_{ll}^{-1}{\bm L}_{lg} & -{\bm D}_{ll}^{-1} {\bm L}_{ll} & {\bm 0}                        \\
        -{\bm M}_{gg}^{-1}{\bm L}_{gg} & -{\bm M}_{gg}^{-1}{\bm L}_{gl}  & -{\bm M}_{gg}^{-1}{\bm D}_{gg}
    \end{bmatrix}.
\end{equation}
Above, ${\bm 0}$ denotes either the matrix or the vector with all components equal to zero,  and $\mathbb{I}$ is the identity matrix, subindices
$g$ and $l$ denote generators and loads respectively.
Inter-area oscillations correspond to the slowest eigenmodes of ${\bm A}$. From Eqs.~\eqref{eq:swinglin} and \eqref{eq:Ax},
they are determined by the Laplacian, the inertia, and the damping matrices.

\subsection{Eigenvectors and Eigenvalues}
Eigenvectors and -values of ${\bm A}$ are easily obtained in the case of homogeneous
damping $d_i=d$ and inertia $m_i=m$.
For each mode, the oscillation frequency reads
\begin{equation}
    \omega_\alpha = \frac{1}{2}\sqrt{\frac{4}{m}\lambda_\alpha -\gamma^2} \, ,
\end{equation}
where $\lambda_\alpha$ is an eigenvalue of ${\bm L}$ and $\gamma = d/m$~\cite{colettaLinearStabilityBraess2016a}. Furthermore, both the angle and frequency components of each eigenmode of ${\bm A}$
are given by those of an eigenmode of ${\bm L}$.
Thus, the oscillation frequency and the mode structure are related to
the eigenvalues and -vectors of ${\bm L}$, when damping and inertia are homogeneous.

\begin{figure}
    \centering
    \includegraphics{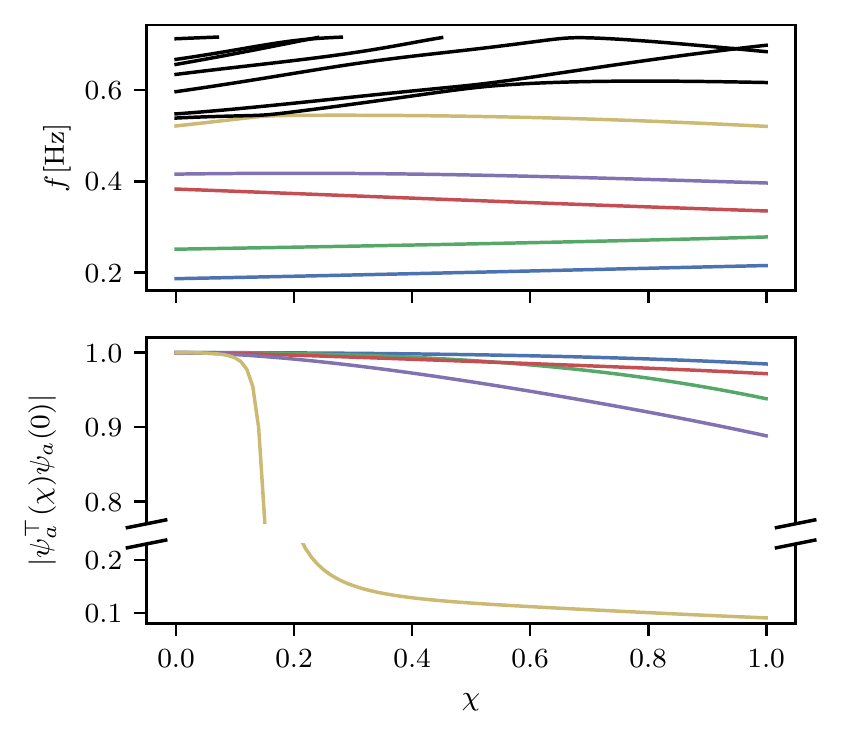}
    \caption{Top panel: oscillation frequencies $f$ of the Kron reduced version of PanTaGruEl vs. the inhomogeneity parameter $\chi$ defined in Eq.~\eqref{eq:inhom}.
        Bottom panel: overlap $|\psi_a^\top(\chi) \psi_a(0)|$ of the five lowest right eigenvectors $\psi_a$ of $\bm A$
        vs. the inhomogeneity parameter $\chi$ defined in Eq.~\eqref{eq:inhom}.
        Colors in both panels refer to the same eigenvectors.}
    \label{fig:chi}
\end{figure}

The homogeneity assumption is not present in real electric power grids, not even after a Kron reduction of the
inertialess load nodes.
A question of central interest is therefore how much do inertia and damping inhomogeneities affect the structure of
the eigenmodes of ${\bm A}$. Our matrix perturbation approach to be presented below shows that long wavelength, slow modes are
only weakly affected by such inhomogeneities. We illustrate this numerically on PanTaGruEl in Fig.~\ref{fig:chi}. We write damping and inertia as
\begin{equation}\label{eq:inhom}
    m_i = \bar{m} + \chi \delta m_i, \qquad d_i = \bar{d} + \chi \delta d_i,
\end{equation}
where the bar indicates the average over the nodes in the network,
$\delta x_i = x_i - \bar{x}$ with the real value $x_i=m_i$ or $d_i$ in the system,
and $\chi \in [0,1]$ is a continuous parameter tuning the system from the homogeneous configuration
($\chi=0$) to its real configuration ($\chi=1$). The top panel of
Fig.~\ref{fig:chi} shows that the oscillation frequencies of the slowest
modes of PanTaGruEl are only weakly sensitive to inertia and damping inhomogeneities.
The overlap $|\psi_\alpha^\top(\chi) \psi_\alpha(0)|$ of the lowest five right eigenvectors at $\chi=0$ and $\chi \in [0,1]$ is next shown in
the bottom panel of
Fig.~\ref{fig:chi}. The data illustrate nicely that
inhomogeneities have only a
minor effect on the oscillation frequency and the spatial structure of the lowest-lying eigenvectors -- those of
interest in this manuscript.

This can be explained by the long-wavelength nature of these modes, which effectively averages
damping and inertia over areas that are so large that the modes do not resolve their inhomogeneities.
Ref.~\cite{pagnierOptimalPlacementInertia2019} found that the change of $\omega_\alpha$ with $\chi$ is to leading order proportional to $\sum_i u_{\alpha i}^2\delta\gamma_i$, where ${\bm u}_\alpha$ is the eigenvector of the Laplacian corresponding to $\lambda_\alpha$ and $\delta\gamma_i = (d_i / m_i - \gamma)$.
For the lowest modes which are approximately constant over large areas, $u_{\alpha i}^2$ can be factored out and $\sum_{i \in \mathrm{area}} \delta\gamma_i \approx 0$, assuming that the distribution of $\delta \gamma_i$ is independent of the geographical location.
A similar argument holds for the eigenvectors.
The corrections to the right eigenvector $\psi_\alpha$ of the stability matrix are given by a sum over all the other eigenvectors $\psi_\beta, \beta \neq \alpha$, each weighted by a factor proportional to $\sum_i u_{\beta i}u_{\alpha i}\delta\gamma_i$.
Assuming again that the lowest modes are mostly constant over large areas, the contributions from the highly fluctuating higher modes average out.

Higher up in the spectrum, modes have a shorter wavelength and hence better geographical resolution, and
these statements obviously break down. This is of no consequence for the validity of our approach to slow, large-wavelength modes.

\section{Matrix Perturbation Theory}\label{sec:pt}

Consider a matrix, which we are able to diagonalize exactly.
Suppose we perturb that matrix as ${\bm L}_0 \rightarrow {\bm L}(\varepsilon)={\bm L}_0 + \varepsilon {\bm L}_1$, with a perturbation matrix
${\bm L}_1$ that does not commute with  ${\bm L}_0$.
Matrix perturbation theory~\cite{hornMatrixAnalysis2012} is a method for describing the change of eigenvalues and -vectors of  ${\bm L}(\varepsilon)$
as a series expansion in the dimensionless parameter $\varepsilon$. It is a standard method of theoretical physics~\cite{sakuraiModernQuantumMechanics2017} that has recently been exported to electric power and other network systems.
It has been used to investigate the change in oscillation frequencies under small changes of certain slow modes in~\cite{maApplication2ndOrder2012}. It has been used in
Ref.~\cite{yangMatrixperturbationtheorybasedOptimalStrategy2018} to construct a control scheme for
the output of generators and enhance power grid stability.
The optimal placement of inertia and damping has further been investigated using perturbation theory in Ref.~\cite{pagnierOptimalPlacementInertia2019}. In a more general context,
Ref.~\cite{yiDiffusionConsensusWeakly2021} uses matrix perturbation theory to investigate a network of networks.

Taken as a whole, perturbation theory is valid as long as $\varepsilon \ll 1$.
However, we argue below that, when applied to a restricted range of low-frequency modes -- such as few of the slowest modes represented in color in Fig.~\ref{fig:chi} -- the validity range generally becomes significantly larger and may even include the $\varepsilon \rightarrow 1$ limit. Before we apply it to our problem, we first
give a brief general description of non-degenerate and degenerate perturbation theory in the next paragraph..

\subsection{General Framework}
\subsubsection{Non-degenerate perturbation theory}
Take a real symmetric matrix ${\bm L}_0$ with known eigenvalues $\lambda_\alpha^{(0)}$ and eigenvectors ${\bm u}_\alpha^{(0)}$.
This matrix is subjected to a perturbation $\varepsilon {\bm L}_1$, where ${\bm L}_1$ is a symmetric matrix which does not
commute with ${\bm L}_0$
and $\varepsilon$ is a dimensionless scalar parameter. Perturbation theory expands
the eigenvalues $\lambda_\alpha$ and -vectors ${\bm u}_\alpha$ of  ${\bm L}(\varepsilon) = {\bm L}_0 + \varepsilon {\bm L}_1$
in a power series in $\varepsilon$,
\begin{subequations}\label{eq:ptgen}
    \begin{align}
        \lambda_\alpha & \approx \lambda_\alpha^{(0)} + \varepsilon \lambda_\alpha^{(1)} + \varepsilon^2 \lambda_\alpha^{(2)} + \mathcal{O}(\varepsilon^3), \\
        {\bm u}_\alpha & \approx {\bm u}_\alpha^{(0)} + \varepsilon {\bm u}_\alpha^{(1)} + \varepsilon^2 {\bm u}_\alpha^{(2)} + \mathcal{O}(\varepsilon^3).
    \end{align}
\end{subequations}
We call $\lambda_\alpha^{(n)}$ (${\bm u}_\alpha^{(n)}$) the $n$th order correction to the eigenvalue (-vector).
The first and second order corrections of the eigenvalues read
\begin{align}
    \lambda_\alpha^{(1)} & = {\bm u}_\alpha^{(0)\top}{\bm L}_1{\bm u}_\alpha^{(0)} \, , \;\;\;\;\;
    \lambda_\alpha^{(2)} = \sum_{\beta\neq \alpha}\frac{\lvert{\bm u}_\beta^{(0)\top}{\bm L}_1{\bm u}_\alpha^{(0)}\rvert^2}{\lambda_\alpha^{(0)}-\lambda_\beta^{(0)}},\label{eq:secorder}
\end{align}
while the first order correction of the eigenvectors reads
\begin{equation}\label{eq:veccor}
    {\bm u}_\alpha^{(1)} = \sum_{\beta\neq \alpha}\frac{{\bm u}_\beta^{(0)\top}{\bm L}_1{\bm u}_\alpha^{(0)}}{\lambda_\alpha^{(0)}-\lambda_\beta^{(0)}}{\bm u}_\beta^{(0)}.
\end{equation}
Higher order corrections can be obtained recursively from the eigenvalue problem
\begin{equation}
    ({\bm L}_0 + \varepsilon{\bm L}_1) {\bm u}_\alpha = \lambda_\alpha {\bm u}_\alpha
\end{equation}
using the series expansion~\eqref{eq:ptgen}. For more details, the reader is referred to Refs.~\cite{Bam20, sakuraiModernQuantumMechanics2017,hornMatrixAnalysis2012}.

\subsubsection{Degenerate perturbation theory}
Eqs.~\eqref{eq:secorder} and \eqref{eq:veccor} are only valid as long as the $\alpha^{\rm{th}}$ unperturbed eigenvalue $\lambda_\alpha^{(0)}$ has
multiplicity one. We call this the nondegenerate case.
Special care needs to be taken when considering corrections to eigenvalues with multiplicity
larger than one. In this case the corresponding eigenvectors are not unique and span the degenerate subspace $D$.
This degenerate subspace has to be considered separately from the rest of the vector space. The eigenbasis spanning $D$
is a priori not uniquely defined, since any normalized linear combination of the degenerate eigenvectors is also an eigenvector.
However there is one and only one linear combination for which the change in eigenvectors is smooth as the
perturbation is turned on.
Degenerate perturbation theory dictates to choose that linear combination as a starting point. It diagonalizes ${\bm L}_1$ within $D$
and is defined by the conditions
\begin{align}\label{eq:firstorder}
    {\bm u}_\alpha^{(0)\top}{\bm L}_1{\bm u}_\beta^{(0)} & = \lambda_\alpha^{(1)}\delta_{\alpha\beta} \;\; \rm{and} \;\;
    {\bm u}_\alpha^{(0)\top}{\bm u}_\beta^{(0)}              = \delta_{\alpha\beta} \, , \forall \alpha\beta \in D \, .
\end{align}
The first condition in Eq.~\eqref{eq:firstorder} readily gives the first-order correction to the degenerate set of eigenvalues.
Higher-order corrections to eigenvalues and eigenvectors of $D$
are given by Eqs.~\eqref{eq:secorder} and \eqref{eq:veccor}, with the substitution
$\beta\neq \alpha$ $\rightarrow$ $\beta\notin D$.

\subsection{Specific Set-up}
We consider a network partitioned into $p$, initially disconnected areas labeled $a$, each containing $n_a$ nodes and
represented by a $n_a \times n_a$ Laplacian matrix ${\bm L}_a$. The
unperturbed Laplacian ${\bm L}_0=\rm{diag}[{\bm L}_a]$ is a block-diagonal matrix.
Because each area is connected, to each block $a$ corresponds
a single eigenvalue $\lambda_a^{(0)} = 0$, associated to an eigenvector being constant in area $a$ and zero everywhere else
\begin{equation}\label{eq:evecarea}
    \tilde{{\bm u}}_a = \frac{1}{\sqrt{n_a}}(0,\dots,0,{\bm 1}_{n_a}^\top,0,\dots,0)^\top \, ,
\end{equation}
where ${\bm 1}$ is the vector of ones, and the tilde means that these eigenvectors do not satisfy~\eqref{eq:firstorder} yet.
From now on, we refer to these eigenvectors as zero-modes. They correspond to each area oscillating coherently on its own, and
we are interested in finding out how they change when the inter-area coupling  is turned on.
If they do not change significantly, the area will engage in coherent oscillations, and we will see that this is the case
for well chosen areas. We therefore focus on the set $D$ spanned by the zero-modes from here on.

The perturbation $\varepsilon {\bm L}_1$  contains the lines that connect the different areas and
$\varepsilon$ tunes the network from having unconnected areas at $\varepsilon = 0$ to recovering the real, fully connected
network at $\varepsilon=1$.
To obtain the linear combination of eigenvectors that satisfy~\eqref{eq:firstorder} we project ${\bm L}_1$ onto $D$ and linearize it.
The projected matrix ${\bm L}_{\mathrm{proj}}$ is given by
\begin{equation}
    \left({\bm L}_{\mathrm{proj}}\right)_{ab} =
    \begin{cases}
        -{\mathcal B}_{ab}/\sqrt{n_an_b} & \quad \text{for } a\neq b \\
        (\sum_c{{\mathcal B}_{ac}})/n_a  & \quad \text{for } a = b
    \end{cases},
\end{equation}
where ${\mathcal B}_{ab}=\sum_{i \in a; j \in b}B_{ij}$ is the sum of all connections between area $a$ and $b$.
${\bm L}_{\mathrm{proj}}$ has a zero eigenvalue with eigenvector
\begin{equation}
    \bm{v}_1 = \left(\sqrt{n_1/N}, \sqrt{n_2/N}, \dots, \sqrt{n_p/N}\right)^\top \, .
\end{equation}
The corresponding linear combination then gives $\sum_a v_{0a}  \tilde{{\bm u}}_a={\bm 1}_N/\sqrt{N}$, i.e., the
global zero-mode of the full network. The $p-1$ other eigenvectors of ${\bm L}_{\mathrm{proj}}$ define $p-1$ other
linear combinations of zero-modes constituting the unperturbed basis in which the perturbation expansions are constructed.
We call these linear combinations "hybridized zero-modes". Our theory to be presented below focuses on them and
on how they evolve as the inter-area connections increase.

The hybridized zero-modes acquire nonzero first-order eigenvalues which are linear in $\varepsilon$, with a slope determined by
\eqref{eq:firstorder}. Second-order corrections to their eigenvalues emerge due to the interaction with non-zero-modes triggered
by $\varepsilon {\bm L}_1$. These corrections read
\begin{equation}
    \lambda_\alpha^{(2)} = -\sum_{\beta \notin D} \frac{\lvert{\bm u}_\beta^{(0)\top}{\bm L}_1{\bm u}_\alpha^{(0)}\rvert^2}{\lambda_\beta^{(0)}} \, .
\end{equation}
Because $\lambda_\alpha^{(0)}=0$ for the hybridized zero-modes, the second-order corrections are in particular negative, reflecting the
eigenvalue repulsion~\cite{Haa18} between the hybridized and the non-zero-modes.
Simultaneously, Gershgorin's circle theorem
guarantees that eigenvalues of  ${\bm L}_0 + \varepsilon{\bm L}_1$ are nonnegative~\cite{hornMatrixAnalysis2012}.
These two effects result in the behavior of the hybridized eigenvalues shown in Fig.~\ref{fig:evalnoxing}.
The eigenvalues of the hybridized modes are shown in color. We see that after a short linear rise captured by first-order perturbation
theory, they all quickly bend downward to reach what looks like a horizontal asymptotic.
Furthermore, only the upper one (dark blue curve) gets close to the
non-zero modes (gray curves) as $\varepsilon$ increases. In the following paragraphs we analyze this behavior in more detail and connect
it to the evolution of the structure of the corresponding modes.

\subsection{Eigenvector mixing and avoided crossings}

\begin{figure}
    \centering
    \includegraphics[width=0.97\columnwidth]{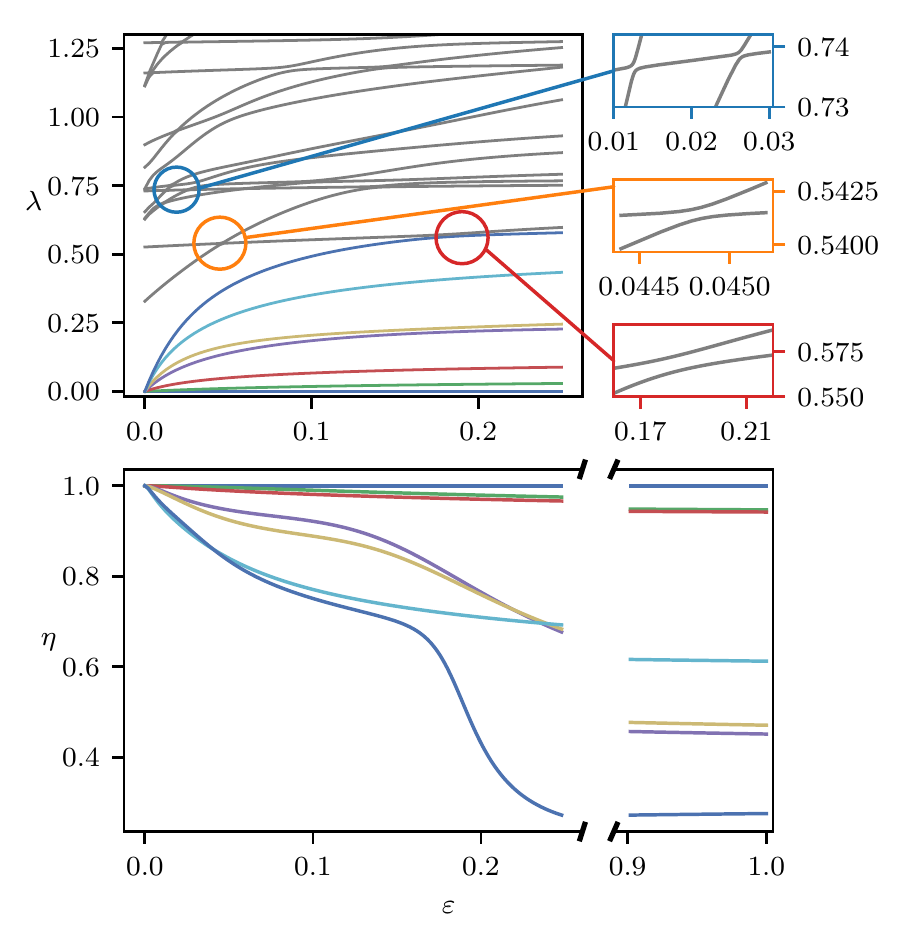}
    \caption{Top panel: Eigenvalues of the Laplacian of the PanTaGruEl model
    initially partitioned into seven areas as a function of the inter-area coupling parameter.
    The global zero-mode and the six hybridized zero-modes are shown in color and several of the lowest non-degenerate
    modes in gray. Circles mark three typical avoided crossings. The
    three right panels make it clear that levels avoid crossing each other. Bottom panel: evolution of
    the scalar product $\eta_\alpha(\varepsilon) = |{\bm u}_\alpha^\top(0) \cdot {\bm u}_\alpha(\varepsilon)|$ of the hybridized zero-modes at
    $\varepsilon = 0$ and at finite $\varepsilon$. The avoided crossing at around $\varepsilon = 0.2$  between the seventh
    (blue) mode and the lowest non-degenerate mode (gray) leads to an abrupt drop in
    $\eta$ for the blue mode. Almost simultaneously, there is an avoided crossing
    between the fourth (violet) and fifth  (beige) eigenvalues, giving a noticeable
    drop in $\eta$ for both modes. The first two hybridized modes (green and red) barely change their structure
    all the way up to $\varepsilon = 1$.}
    \label{fig:evalnoxing}
\end{figure}

Our conjecture is that inter-area oscillations directly
originate from the set of zero-modes just described, corresponding to an appropriately chosen
network partition. One key point is to show that, at least for few of the lowest-lying
hybridized zero-modes diagonalizing the projection
of ${\bm L}_1$ onto the degenerate zero-subspace $D$, the linear combination is only weakly sensitive to the connection
parameter $\varepsilon \in [0,1]$, i.e., well beyond the expected validity range of both perturbation theory and the standard theory of
inter-area oscillations. To qualitatively
understand why that is so, we first
recall that, unless some underlying symmetry is present, the eigenvalues of a parameter-dependent matrix such as ${\bm L}(\varepsilon)$
generically do not cross as $\varepsilon$ is varied. This resistance of eigenvalues to crossings is illustrated in Fig.~\ref{fig:evalnoxing}
which shows
that eigenvalues of ${\bm L}(\varepsilon)$ for a seven-area partition of the PanTaGruEl model exhibit avoided crossings~\cite{Haa18} -- they can get very close to one another
but generically do not cross.
To further demonstrate this, zoom-ins on three characteristic avoided crossings are shown in the top right panels of Fig.~\ref{fig:evalnoxing}.
Second, we recall von Neumann and Wigner's argument that an eigenvector of a parameter-dependent matrix
remains essentially the same as this parameter is varied, as long as its eigenvalue stays away from avoided crossings~\cite{vonneumannUberVerhaltenEigenwerten1929}. We conclude then that, if an eigenvalue ${\bm L}(\varepsilon)$
does not go through an avoided
crossing as $\varepsilon$ increases from 0 to 1, the structure of the corresponding eigenvector does not change much

This argument is corroborated by the data in Fig.~\ref{fig:evalnoxing}. The change in eigenvector structure can be
measured by the overlap $\eta_\alpha(\varepsilon) = \lvert{\bm u}_\alpha^\top(0) \cdot {\bm u}_\alpha(\varepsilon)\rvert$ of an eigenvector
at $\varepsilon = 0$ and at finite $\varepsilon$.
Fig.~\ref{fig:evalnoxing} clearly shows that the lowest two hybridized modes (green and red curves) do not change their structure.
The effect of an avoided crossing on eigenmode structure is clearly illustrated in Fig.~\ref{fig:evalnoxing}. In the top panel, one sees that the seventh hybridized (blue curve) and first non-hybridized eigenmodes get close to each other
at around $\varepsilon \simeq 0.18$ (indicated by the red circle), but do not cross. Simultaneously, a drop in $\eta$ for the  seventh hybridized eigenmode is
observed in the bottom panel at the same value of $\varepsilon$. This reflects a scrambling of the eigenmode due to the avoided crossing. The same behavior is observed for the fourth (violet) and fifth (beige) hybridized modes, coincidentally at about the same value of $\varepsilon$.

The occurrence of avoided crossings as $\varepsilon$ increases signals the onset of eigenvector mixing, beyond which
the  eigenvectors of ${\bm L}(0)$ are no longer representative of the eigenvectors of ${\bm L}(\varepsilon)$.
Before we discuss such occurrences, we first derive a general criterion for the breakdown of perturbation theory for eigenvalues.

\subsection{Perturbation Series Convergence Criteria}

According to d'Alembert's ratio criterion the convergence radius $r$ of a series $f(x) = \sum_{n=0}^{\infty} c_n x^n$ is given by $r = \lim_{n\to\infty}\lvert \frac{c_n}{c_{n+1}}\rvert.$
For the perturbation expansion of Eqs.~\eqref{eq:ptgen} to converge up to $\varepsilon\to1$, the corrections
need to be smaller with each order. For the eigenvalue expansion this translates into
\begin{equation}
    \left\lvert\frac{\lambda_\alpha^{(k+1)}}{\lambda_\alpha^{(k)}}\right\rvert < 1.
\end{equation}
We approximate the expression on the left-hand side, assuming that
${\bm u}_\alpha^{(0)\top}{\bm L}_1{\bm u}_\beta^{(0)}\simeq \frac{1}{N-p}\sum_\gamma {\bm u}_\alpha {\bm L}_1{\bm u}_\gamma$
is close to its average value for $\alpha \in D$ and for all $\beta \notin D$. Under this assumption, Eq.~\eqref{eq:secorder} reads
\begin{equation}
    \lambda_\alpha^{(2)} \approx -\bar{S}_\alpha \sum_{\beta\notin D} \frac{1}{\lambda_\beta^{(0)}},
\end{equation}
with
\begin{equation}
    \bar{S}_\alpha = \frac{1}{N-p}\sum_{\beta\notin D}\lvert{\bm u}_\beta^{(0)\top}{\bm L}_1{\bm u}_\alpha^{(0)}\rvert^2.
\end{equation}
We can get rid of the sum over $\beta$ using
\begin{equation}
    \sum_{\alpha\notin D}{\bm u}_\alpha^{(0)}{\bm u}_\alpha^{(0)\top} = \mathbb{I} - \sum_{\alpha\in D}{\bm u}_\alpha^{(0)}{\bm u}_\alpha^{(0)\top} \, ,
\end{equation}
from which we obtain, with a little bit of algebra,
\begin{equation}
    \begin{split}
        \bar{S}_\alpha &=
        \frac{1}{N-p}\left({\bm u}_\alpha^{(0)\top}{\bm L}_1^2{\bm u}_\alpha^{(0)} - (\lambda_\alpha^{(1)})^2\right).
    \end{split}
\end{equation}
This expression is helpful, because it no longer contains a sum over $\beta \notin D$ and expresses $\lambda_\alpha^{(2)}$
only as a function of $\lambda_\alpha^{(1)}$ and the expectation value of  the squared  interaction Laplacian over the eigenvector
${\bm u}_\alpha^{(0)}$.
Note that the latter, despite fulfilling~\eqref{eq:firstorder} are not eigenvectors of ${\bm L}_1$, so that  ${\bm u}_\alpha^\top{\bm L}_1^2{\bm u}_\alpha\neq(\lambda_\alpha^{(1)})^2$.
The second order correction finally becomes
\begin{equation}
    \lambda_\alpha^{(2)} \approx -\bar{S}_\alpha\sum_{\beta\notin D}\frac{1}{\lambda_\beta^{(0)}} = -\bar{S}_\alpha\sum_{a=1}^p\frac{1}{n_a}\mathrm{Kf}_1^{(a)},
\end{equation}
where we introduced the generalized Kirchhoff indices of the disconnected subgraphs~\cite{tylooRobustnessSynchronyComplex2018},
\begin{equation}
    \mathrm{Kf}_m^{(a)} = n_a\sum_{\substack{\alpha \in \text{area } a\\\alpha \notin D}}\left(\lambda_\alpha^{(0)}\right)^{-m} \, .
\end{equation}

From this analysis we find the following criterion for convergence
\begin{equation}
    \left\lvert\frac{\lambda_\alpha^{(2)}}{\lambda_\alpha^{(1)}}\right\rvert = \left\lvert\frac{\bar{S}_\alpha}{\lambda_\alpha^{(1)}}\sum_{a=1}^p\frac{1}{n_a}\mathrm{Kf}_1^{(a)}\right\rvert< 1\label{eq:convfirst}.
\end{equation}
The criterion combines an overall network characteristic with a mode-specific characteristic.
To be satisfied, Eq.~\eqref{eq:convfirst} requires that either (i) $\sum_{a=1}^p\frac{1}{n_a}\mathrm{Kf}_1^{(a)}\ll 1$  or (ii) $\bar{S}_\alpha/\lambda_\alpha^{(1)} \ll 1$, or both.
The first condition, that the weighted sum of subgraph Kirchhoff indices is small, requires that intraconnections are strong within
subgraphs. This is so, because the Kirchhoff index is the average resistance distance in the graph.
This first condition is consistent with recent works which find that the coherence of networks increases with the first non-zero eigenvalue of the network Laplacian, even for higher-order generator models~\cite{Min_2019}.
The second condition requires that the inter-area connection strength felt by the $\alpha^{\rm th}$ hybridized mode is small. In this sense,
Eq.~\eqref{eq:convfirst}
is qualitatively similar to the conditions for  validity of singular perturbation theory and time-scale separation~\cite{chowSingularPerturbationAnalysis1978,chowTimeScaleModelingDynamic1982,chowPowerSystemCoherency2013,romeresNovelResultsSlow2013,dateAggregationPropertiesLinearized1991}. There is however a significant difference in that the condition of
Eq.~\eqref{eq:convfirst} applies to each hybridized zero-mode individually. In particular, perturbation theory may capture certain modes
efficiently, while failing for others. This difference is of key importance, as we will see that for
modes at the bottom edge of the spectrum -- corresponding to the low frequency inter-area oscillations -- perturbation theory remains valid at higher $\varepsilon$, whereas the theory breaks down earlier for
other modes higher up in the spectrum. The standard criteria for validity of
singular perturbation theory and time-scale separation
are global, therefore they implicitly request that the theory is applicable to all states.
They are therefore too restrictive.

The criterion of Eq.~\eqref{eq:convfirst} is helpful; however, it still needs to be computed numerically. It shows that
the breakdown of perturbative approaches is mode-specific.
Similar criteria can be found for higher orders of perturbation theory. Even though we cannot evaluate d'Alembert's criterion for $n\to\infty$ we find that the first few orders already give a good approximation of the convergence.

\subsection{Avoided Crossings}
The criteria for validity of perturbation theory derived in the previous section are mode-dependent. They raise the important issue
of determining which modes are best captured by perturbation theory for larger $\varepsilon$.
To that end we recall what is sometimes referred to as the von Neumann-Wigner theorem~\cite{vonneumannUberVerhaltenEigenwerten1929}, which states in our case that, as $\varepsilon$ varies, pairs of eigenvalues of ${\bm L}(\varepsilon) = {\bm L}_0 + \varepsilon {\bm L}_1$ may undergo close encounters, however, they will generally avoid crossing each other's path. Von Neumann and Wigner further argued that it is at these avoided crossings
that eigenvectors get mixed and that their structure changes fundamentally.
Conversely, as long as an eigenmode is not undergoing any avoided crossing, then  its structure does not change much.
Therefore, predicting the first occurrence of an avoided crossing among the lowest eigenvectors is key to understand how far
$\varepsilon$ can grow, without altering the structure of an eigenmode. Stated otherwise, if the first few hybridized zero-modes do not
undergo any avoided crossing until $\varepsilon=1$, then the initial area partition at $\varepsilon=0$ predicts the
inter-area mode structure at $\varepsilon=1$ with great precision.
We analyze the situation for the case of two and of more than two areas.

\subsubsection{Case of two areas}

With two areas, there is one global zero-mode and one hybridized zero-mode. We want to derive a condition for the eigenvalue of the
latter not to undergo an avoided crossing with the third eigenvalue, corresponding to the lowest non-zero-mode.
To that end we first show that the first order correction $\lambda_2^{(1)}$ to the hybridized zero-mode
gives an upper bound to its true eigenvalue.
Consider the first non-zero eigenvalue $\lambda_2(\varepsilon)$ at position $\tilde{\varepsilon}$ and $\tilde{\varepsilon} + \delta\tilde{\varepsilon}$ where $0<\delta\tilde{\varepsilon} \ll 1$.
The slope at both points is given by the first order perturbation theory
\begin{subequations}
    \begin{align}
        \diff.pvrule.{\lambda_2}{\varepsilon}[\tilde{\varepsilon}]                             & = {\bm u}_2(\tilde{\varepsilon})^\top{\bm L}_1{\bm u}_2(\tilde{\varepsilon}),                                                                           \\
        \diff.pvrule.{\lambda_2}{\varepsilon}[\tilde{\varepsilon} + \delta\tilde{\varepsilon}] & = {\bm u}_2(\tilde{\varepsilon} + \delta\tilde{\varepsilon})^\top{\bm L}_1{\bm u}_2(\tilde{\varepsilon} + \delta\tilde{\varepsilon}).\label{eq:derived}
    \end{align}
\end{subequations}
The eigenvector at $\tilde{\varepsilon} + \delta\tilde{\varepsilon}$ can be expressed in terms of the eigenvectors at $\tilde{\varepsilon}$ using~\eqref{eq:veccor}
\begin{equation}
    {\bm u}_2(\tilde{\varepsilon} + \delta\tilde{\varepsilon}) = {\bm u}_2(\tilde{\varepsilon}) + \delta\tilde{\varepsilon}\sum_{\alpha > 2}\frac{{\bm u}_\alpha(\tilde{\varepsilon}){\bm L}_1{\bm u}_2(\tilde{\varepsilon})}{\lambda_2(\tilde{\varepsilon}) - \lambda_\alpha(\tilde{\varepsilon})}{\bm u}_\alpha + \mathcal{O}(\delta\tilde{\varepsilon}^2)
\end{equation}
Eq.~\eqref{eq:derived} then becomes
\begin{equation}\label{eq:shape}
    \begin{split}
        \diff.pvrule.{\lambda_2}{\varepsilon}[\tilde{\varepsilon} + \delta\tilde{\varepsilon}]=&{\bm u}_2(\tilde{\varepsilon})^\top{\bm L}_1{\bm u}_2(\tilde{\varepsilon})\\
        &+2\delta\tilde{\varepsilon}\sum_{\alpha > 2}\frac{\left\lvert{\bm u}_\alpha(\tilde{\varepsilon}){\bm L}_1{\bm u}_2(\tilde{\varepsilon})\right\rvert^2}{\lambda_2(\tilde{\varepsilon}) - \lambda_\alpha(\tilde{\varepsilon})} + \mathcal{O}(\delta\tilde{\varepsilon}^2).
    \end{split}
\end{equation}
The second term on the right-hand side of Eq.~\eqref{eq:shape} is always negative, because the denominator of each term is negative
while the numerator is positive.
We therefore conclude
\begin{equation}\label{eq:concave}
    0< \diff.pvrule.{\lambda_2}{\varepsilon}[\tilde{\varepsilon} + \delta\tilde{\varepsilon}] \le \diff.pvrule.{\lambda_2}{\varepsilon}[\tilde{\varepsilon}],
\end{equation}
where the lower bound is due to ${\bm L_1}$ being Laplacian.
Thus, $\lambda_2(\varepsilon)$ is concave and it is upper bounded by its first order perturbation theory correction
at $\varepsilon=0$.
Note that Eq.~\eqref{eq:concave} remains valid, regardless of the number of areas.

Second, we derive a lower limit for the third smallest eigenvalue. Weyl's theorem~\cite{hornMatrixAnalysis2012} for the eigenvalues of the sum of two symmetric matrices states that
\begin{equation}
    \lambda_\alpha(A + B) \ge \lambda_\alpha(A) + \lambda_1(B).
\end{equation}
With $A={\bm L_0}$ and $B= \varepsilon {\bm L_1}$, this gives
$\lambda_3(\varepsilon) \ge \lambda_3(\varepsilon = 0)$.
Thus, a sufficient condition that there is no avoided crossing between $\lambda_2$ and $\lambda_3$ up to $\varepsilon=1$ is
\begin{equation}\label{eq:2areacondition}
    {\bm u}_2(\varepsilon=0)^\top{\bm L}_1{\bm u}_2(\varepsilon=0) < \lambda_3({\bm L}_1).
\end{equation}
This condition underestimates the validity of perturbation theory.

\subsubsection{Case of more than two areas}

When there are more than two areas, hybridized zero-modes may interact with one another via avoided crossings of their respective
eigenvalues. The occurrence of these crossings is harder to predict than those between a single hybridized mode and the first non-zero mode
treated in the previous paragraph. Here we focus on the occurrence of an avoided crossing between the first and second hybridized zero-mode.
Eq.~\eqref{eq:concave} remains valid and $\lambda_2$ is a concave and monotonously increasing function of $\varepsilon$. This behavior is captured only at small enough values of $\varepsilon$ by a truncated perturbative series.
This is not too big a restriction, since one expects avoided crossings to occur at low values of $\varepsilon$, because it is there that
the largest changes to the topology of the network occur -- going from unconnected to connected. Here we
restrict our discussions to series
truncated at (and including) second order in $\varepsilon$ and construct conditions under which there is no avoided crossing between
$\lambda_2$ and $\lambda_3$ before $\varepsilon_{\rm max}= -\lambda_2^{(1)}/2 \lambda_2^{(2)}$, where the truncated series reaches its maximum.

There are then two separate conditions under which there is no avoided crossing between $\lambda_2$ and $\lambda_3$. The first one
is when $\lambda_2^{(2)} < \lambda_3^{(2)}$, because then second order corrections increase the already increasing distance between
$\lambda_2$ and $\lambda_3$ in first order perturbation theory. The second one is when the second order corrections are not sufficient
to induce a crossing between the series for $\lambda_2$ and $\lambda_3$ truncated at second order before $\varepsilon_{\rm max}$.
These two conditions read
\begin{eqnarray}\label{eq:mulareas}
    \quad\frac{\lambda_2^{(1)}-\lambda_3^{(1)}}{\lambda_3^{(2)} - \lambda_2^{(2)}}&<&0 \,,  \;\; \rm{and} \;\;
    \frac{\lambda_2^{(1)}-\lambda_3^{(1)}}{\lambda_3^{(2)} - \lambda_2^{(2)}} >  a \, \varepsilon_{\rm max} \, .  \quad
\end{eqnarray}
While the argument above leads to $a=1$, we found numerically that a value of $a=1.5$ gives better predictions.

\section{Numerical validation}\label{sec:results}
We  validate the perturbation theory presented above by numerical investigations on three different networks:
(i) a synthetic two area network, (ii) the IEEE RTS 96 test system, and (iii) the PanTaGruEl model of the synchronous grid of continental Europe.

\subsection{Synthetic Two Area Network}

\begin{figure}
    \centering
    \includegraphics[width=\columnwidth]{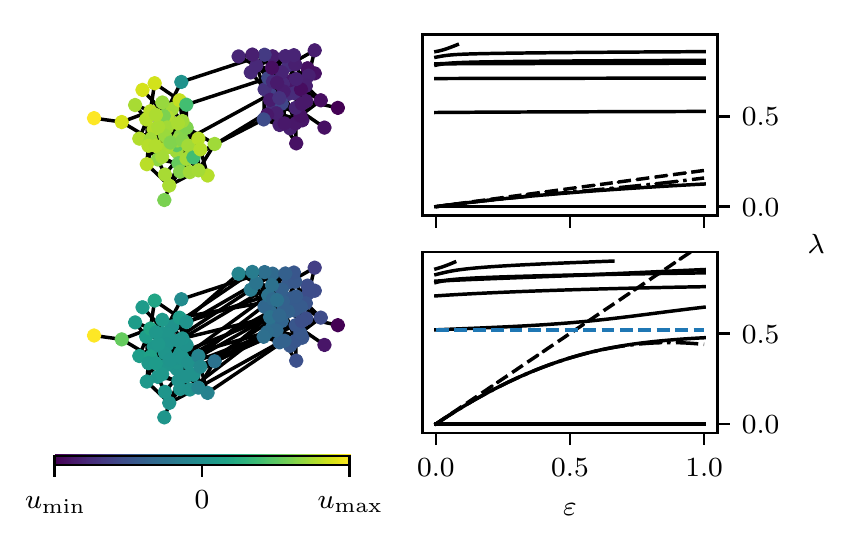}
    \caption{Left column: Synthetic two area network  with five (top) and twenty-five (bottom) inter-area connection lines. In both cases,
        the lowest eigenmode at $\varepsilon=1$ is color-coded on the network nodes. Right column: evolution of the spectrum as the connecting
        line capacities increase. The eigenvalues of ${\bm L}(\varepsilon)$ are shown as solid lines, the first order perturbation corrections as black dashed lines, and the corrections up to third order as dotted-dashed lines. $\lambda_3(\varepsilon = 0)$ is shown as a dashed blue line. For five connections (top), there is no avoided crossing in the spectrum, accordingly, the structure of the lowest eigenmode reflects the two network areas. This can be seen by the eigenmode being almost constant on each area (positive on the left, light-green area and negative on the right, dark-purple area). For twenty-five connections (bottom) the situation is clearly different, which is due to the presence
        of an avoided crossing between the first and second non-zero modes around $\varepsilon \simeq 0.7$ (observable by the bending of the second and third eigenvalues in the bottom-right panel).
        This dramatically changes the structure of the eigenmode which no longer resolves the two areas. The avoided crossing
        is predicted by first-order perturbation theory (crossing of the dashed lines).
    }
    \label{fig:synthgraph}
\end{figure}

We generate two Erdős-Rényi graphs, each with 50 nodes and a connection probability of $p=0.1$, resulting in each node being
connected to a bit less than 5 other nodes on average. All lines in these graphs have the same
capacity, $B_{ij}\equiv1$, and we next connect them via (i) 5 and (ii) 25 lines, each with the same capacity $B_{ij}\equiv \varepsilon \in [0,1]$.
The networks are shown in Fig.~\ref{fig:synthgraph}. The additional connections change the number of lines
from 228 to 233 in the first case and to 253 in the second case. In both instances, two areas are still clearly defined, yet in the second
case, we will see that the occurrence of an avoided crossing as $\varepsilon$ increases totally mixes the structure of the single hybridized
zero mode in this two-area set-up.

In the first case, we introduce five random connections between the two areas. We find that the left-hand side in Eq.~\eqref{eq:convfirst} is significantly smaller than one, so that perturbation theory should be valid and the slowest inter-area mode should reflect the structure of the network.
This analysis is confirmed by calculating the actual perturbational corrections up to third order (not discussed above, for details, see Ref.~\cite{hornMatrixAnalysis2012})
and noticing that with each order the approximation converges to larger $\varepsilon$.
We furthermore check that Eq.~\eqref{eq:2areacondition} is satisfied, therefore we expect that the first eigenmode is well captured.
This is the case with $\eta(\varepsilon=1) = 0.99$ indicating that the eigenvector barely changes with increasing $\varepsilon$. The top right panel in Fig.~\ref{fig:synthgraph}
shows that in this case, $\lambda_2(\varepsilon)$ undergoes no avoided crossing.

In the second case, twenty-five random connections are introduced between the two areas, so that
every other node has a connection to the other area on average. The bottom right panel in
Fig.~\ref{fig:synthgraph} shows that, again in this case,
perturbation theory approximates the eigenvalue well even for $\varepsilon \to 1$. However,
this time we find that Eq.~\eqref{eq:2areacondition} is not met. We therefore expect the lowest eigenmode to undergo an avoided crossing,
and this is confirmed in the bottom right panel of Fig.~\ref{fig:synthgraph}, where an avoided crossing is visible at $\varepsilon \approx 0.7$.

\subsection{IEEE RTS 96 test system}
\begin{figure}
    \centering
    \includegraphics[width=\columnwidth]{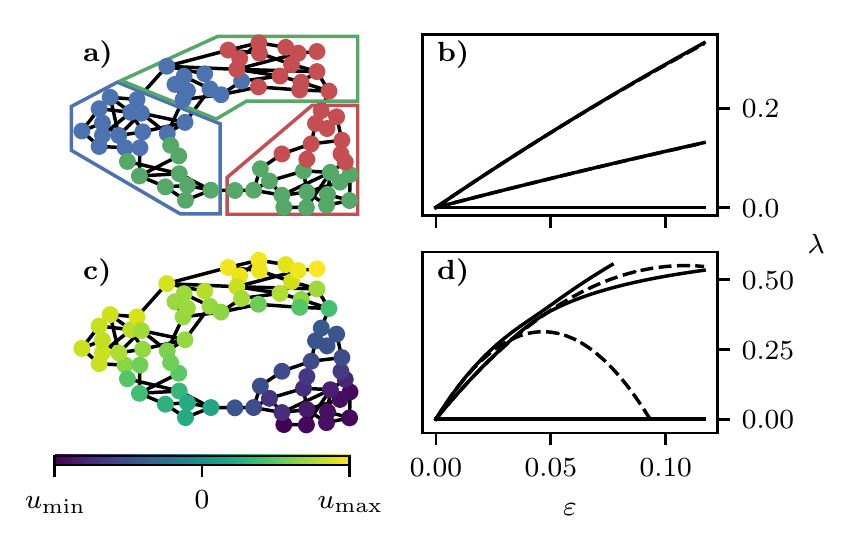}
    \caption{a) IEEE RTS 96 test system, with the obvious three-area aggregation (red, blue and green polygons), and a
    counterintuitive aggregation (red, blue and green nodes). b) $\varepsilon$-dependence of the three lowest eigenvalues
    for the correct aggregation. There is no avoided crossing and the eigenmode preserve their structure all the way to
    $\varepsilon=1$. c) Color-coded structure of the lowest mode of the Laplacian for $\varepsilon=1$.
    Its structure is already well predicted by our first-order perturbation theory with the correct aggregation,
    giving an overlap $\eta_2(\varepsilon) = |{\bm u}_2^\top(\varepsilon=0) \cdot {\bm u}_2(\varepsilon=1)| \simeq 0.96$.
    d) $\varepsilon$-dependence of the three lowest eigenvalues
    for the counterintuitive aggregation. There is an avoided crossing between the second and the third eigenvalues
    and their structure is significantly changed well before
    $\varepsilon=1$. The second order corrections are shown as dashed lines to visualize Eq.~\eqref{eq:mulareas}.\label{fig:rtstopo}}
\end{figure}

The IEEE RTS 96 test system consists of 73 nodes divided into three well-defined areas as shown in Fig.~\ref{fig:rtstopo}~\cite{griggIEEEReliabilityTest1999}. We consider two different initial aggregations, first along the obvious area boundaries, second
along three lines cutting across each area.

In the first case, we find that the
slow eigenvalues are well captured by perturbation theory, and that moreover there is no avoided crossing affecting the two lowest
non-zero eigenvalues corresponding to the hybridized zero-modes. Accordingly, the corresponding modes retain the structure of the
initial aggregation.

In the second case, the chosen initial aggregation results in large perturbative corrections and eventually to the lowest eigenvalues
undergoing an avoided crossing, as predicted by Eq.~\eqref{eq:mulareas}.
The avoided crossing is shown in the bottom right panel of Fig.~\ref{fig:rtstopo}. This shows that
an incorrect initial aggregation leads to avoided crossings, which are the mechanism for mode mixing.
Numerical investigations show that indeed the structure of both eigenvectors arising from the degenerate subspace changes almost completely ($\eta(\varepsilon=1) < 0.06$).

\subsection{PanTaGruEl}\label{sec:pantagruel}

The PanTaGruEl model of the synchronous grid of continental Europe
consists of 3809 nodes connected by 4944 power lines
~\cite{pagnierOptimalPlacementInertia2019,pagnierInertiaLocationSlow2019}. With the dispatch used for this paper,
there are 468 generators. To illustrate the validity
of the theory presented above, we used the standard aggregation algorithm of Ref.~\cite{chowPowerSystemCoherency2013}.
We found that, to capture the inter-area oscillations, an aggregation into seven areas works well, and that the number and structure of these
modes does not change as the number of areas increases.

\begin{figure}[htbp]
    \centering
    \includegraphics[width=\columnwidth]{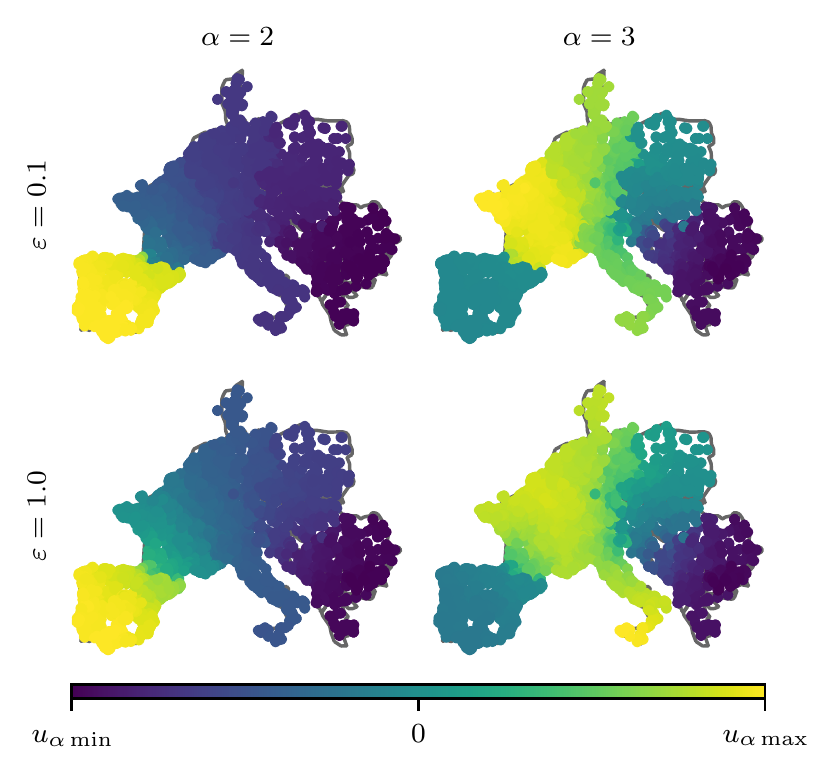}
    \caption{Structure of the first ($\alpha=2$) and the second ($\alpha=3$) hybridized zero-mode of \emph{PanTaGruEl} in the weakly (top panel) and fully (bottom) connected cases. The colors correspond to the value of the eigenvector $u_{\alpha,i}$ on the corresponding node $i$. The mode structure remains the same as $\varepsilon$ goes to $\varepsilon=1$.}
    \label{fig:eigvecspanta}
\end{figure}

PanTaGruEl is a strongly connected network with no obvious area separation (except perhaps the Iberian Peninsula) and,
not surprisingly,
the convergence criterion~\eqref{eq:convfirst} is by far not met. Yet, the slowest eigenvectors are not
much affected by the inter-area connections.
This is predicted by the criteria of Eqs.~\eqref{eq:mulareas}, which are met, and corroborated by numerical data.
First, Fig.~\ref{fig:eigvecspanta} shows the lowest nonzero eigenvector of the network Laplacian
at $\varepsilon=0.1$ and $\varepsilon=1$. Clearly the general structure remains the same, regardless of the inter-area coupling strength.
This is corroborated by the overlap data shown in Fig.~\ref{fig:evalnoxing}.

Interesting is the behavior of the
fourth and fifth eigenvectors which undergo an avoided crossing shown in Fig.~\ref{fig:evalnoxing} at $\varepsilon \simeq 0.2$.
Their behavior illustrates the eigenvector mixing process discussed above, as after the avoided crossing, $\varepsilon > 0.2$,
the actual eigenvectors are given by ${\bm u}_{4,5} \approx ({\bm \tilde{u}}_4 \pm {\bm \tilde{u}}_5)/\sqrt{2}$,
where ${\bm \tilde{u}}_i$ denotes the eigenvector before the avoided crossing. This
results in overlaps $\eta_{4,5}(\varepsilon=1) \approx 0.45$ at $\varepsilon=1$ for both eigenvectors. The seventh eigenvector also undergoes an avoided
crossing at $\varepsilon \simeq 0.2$, which mixes it with the
high-lying eigenvectors. From Fig.~\ref{fig:evalnoxing} we see that this is accompanied by an abrupt decrease of $\eta_7$.
These phenomena nicely illustrate the direct connection between avoided crossings and changes in the eigenvectors structure.
\begin{figure}
    \centering
    \includegraphics{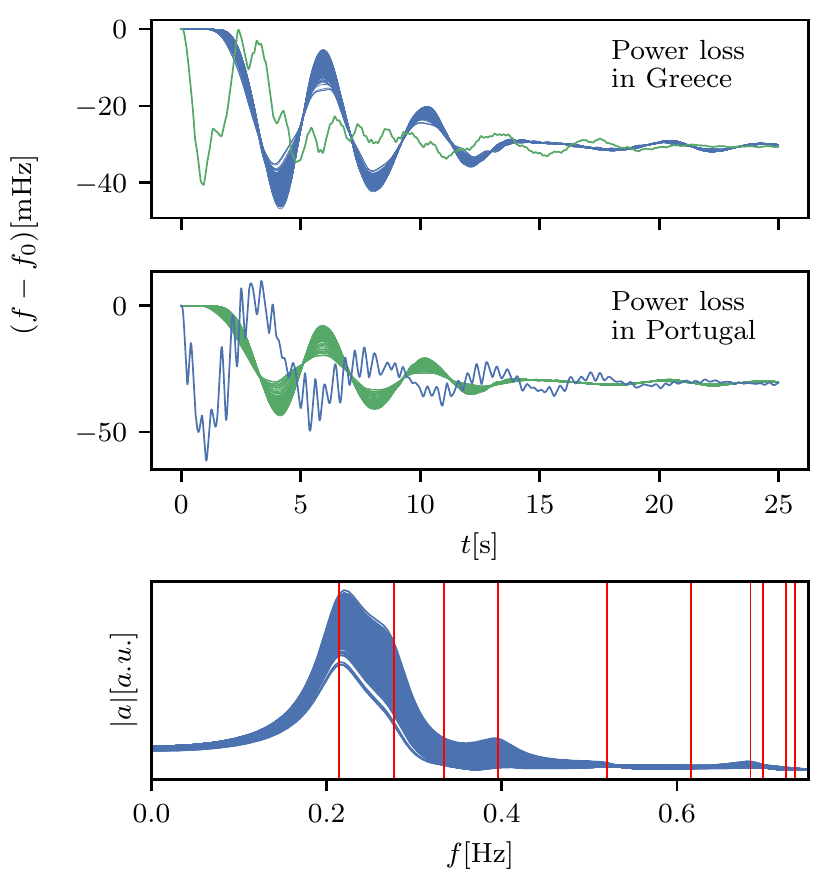}
    \caption{Top and middle: frequency response of the 368 nodes in the Balkan area (green) and the 981 nodes in the Iberian Peninsula (blue) to a 900~MW power loss in the opposing area. The faults occur at the generators indicated by crosses in Fig~\ref{fig:noise}. The frequency response of a generator in the corresponding perturbed area is shown in each panel to demonstrate the inter-area character. Bottom: Fourier transform of the frequency response in the Iberian Peninsula with the average removed. The red lines indicate the eigenfrequencies of the full system. It is visible that the inter-area oscillations are mainly carried by the two lowest frequencies.}
    \label{fig:fault}
\end{figure}

Figs.~\ref{fig:noise} and \ref{fig:eigvecspanta} show
that the lowest hybridized zero-mode essentially resides in the Iberian Peninsula and in the Balkans.
These are two of the seven areas in our aggregation. We found that oscillations between these two areas are triggered
by a perturbation in either one. This is shown in Fig.~\ref{fig:noise} in the case of a noisy power injection. The chosen noise is an Ornstein-Uhlenbeck
noise with a correlation time much larger than the oscillation frequency, and we found that the latter is close to the oscillation frequencies
of the two lowest hybridized zero-modes and not related to any perturbation time scale. These two modes are excited by the perturbation, and the addition of their components
evidently resolves the two areas as shown in Fig.~\ref{fig:noise}.

As another example, we finally investigate the reaction of the system to a 900~MW power loss.
Fig.~\ref{fig:fault} shows the response of the Balkan area and the Iberian Peninsula to a fault in the opposing area.
As before, all nodes within one area respond coherently -- with the same frequency and phase -- to a fault in the opposing area, with only moderate variations of the response amplitude.
The Fourier transform of the oscillation response in the Iberian Peninsula is further shown in the bottom panel. Superimposed on it are the locations of the eigenfrequencies of the $\bm A$-matrix, and the Fourier spectrum
indicates that only the first two eigenmodes are excited, with a broadening originating from the damping in Eq.~\eqref{eq:swinglin}. This confirms our above claim that
inter-area oscillations in the PanTaGruEl network are mainly carried by the two lowest modes of the system.

\section{Conclusion}\label{sec:conclusions}

The standard approach to slow coherency successfully predicts the structure of
slowly oscillating, long-wavelength inter-area modes, even in strongly connected power networks that lie outside its range of validity.
The theory presented above puts this theory on solid grounds even in such well connected networks. Our line of reasoning goes as follows.

First, we recalled that in homogeneous systems with constant inertia and damping, small-signal oscillations in the swing equations \eqref{eq:generators}
are carried by eigenmodes of the ${\bm A}$-matrix of Eq.~\eqref{eq:Ax}.
The structure of these modes is solely determined by the eigenmodes of the Laplacian matrix of the
network~\cite{colettaLinearStabilityBraess2016a}.

Second, when dealing with slow coherency/inter-area oscillations, the focus is on the slowly oscillating modes. A recent
extension of Courant's nodal domain theorem~\cite{Cou23} shows that the slow modes have large nodal domains~\cite{Urs18}.
Because of that, these slow modes are only poorly resolving inhomogeneities in inertia and damping and are therefore
much less sensitive to them. Perturbation theory makes this quantitative and shows that, while in the presence
of significant inhomogeneities, most eigenmodes of the ${\bm A}$-matrix acquire a structure not necessarily captured by those of the
network Laplacian ${\bm L}$.
The long-wavelength modes of ${\bm A}$ retain the structure of the slow modes of ${\bm L}$.

Third, using perturbation theory, we showed that an appropriate choice of disaggregation of the network into initially disconnected
areas captures the slow modes from the hybridization of the zero modes of each area Laplacian, even when the inter-area couplings
are restored. This clarifies the origin of the slow coherency, inter-area modes, and in particular explains how they often are quasi-homogeneous
over large areas -- they originate from area-Laplacian zero-modes that are exactly constant there.

The aim of our theory was not to identify
the optimal initial aggregation. We found numerically that standard numerical aggregation algorithms do a good job at predicting that. Our theory
fills an important theoretical gap, in that it explains (i) the origin of the slow inter-area modes and (ii) why the standard approach to
inter-area oscillations still works well outside its range of validity. Doing so, we closed a number of loopholes
in the theory of inter-area oscillations in transmission power networks.

\section*{Acknowledgment}
We are grateful to M. Tyloo for discussions at the early stage of this project and to G. Evequoz for discussions on Courant's nodal domain theorem.


\begin{thebibliography}{10}
    \providecommand{\url}[1]{#1}
    \csname url@samestyle\endcsname
    \providecommand{\newblock}{\relax}
    \providecommand{\bibinfo}[2]{#2}
    \providecommand{\BIBentrySTDinterwordspacing}{\spaceskip=0pt\relax}
    \providecommand{\BIBentryALTinterwordstretchfactor}{4}
    \providecommand{\BIBentryALTinterwordspacing}{\spaceskip=\fontdimen2\font plus
        \BIBentryALTinterwordstretchfactor\fontdimen3\font minus
        \fontdimen4\font\relax}
    \providecommand{\BIBforeignlanguage}[2]{{%
                \expandafter\ifx\csname l@#1\endcsname\relax
                    \typeout{** WARNING: IEEEtran.bst: No hyphenation pattern has been}%
                    \typeout{** loaded for the language `#1'. Using the pattern for}%
                    \typeout{** the default language instead.}%
                \else
                    \language=\csname l@#1\endcsname
                \fi
                #2}}
    \providecommand{\BIBdecl}{\relax}
    \BIBdecl

    \bibitem{rogersPowerSystemOscillations2000}
    G.~Rogers, \emph{Power System Oscillations}.\hskip 1em plus 0.5em minus
    0.4em\relax {Boston, MA}: {Springer US}, 2000.

    \bibitem{venkatasubramanianAnalysis1996Western2004}
    V.~Venkatasubramanian and Y.~Li, ``Analysis of 1996 western american
    blackouts,'' in \emph{Bulk {{Power System Dynamics}} and {{Control}} -
    {{VI}}}, {Cortina d'Ampezzo}, Aug. 2004, pp. 685--721.

    \bibitem{Zha16}
    \BIBentryALTinterwordspacing
    X.~Zhang, C.~Lu, S.~Liu, and X.~Wang, ``A review on wide-area damping control
    to restrain inter-area low frequency oscillation for large-scale power
    systems with increasing renewable generation,'' \emph{Renewable and
        Sustainable Energy Reviews}, vol.~57, pp. 45--58, 2016. [Online]. Available:
    \url{https://www.sciencedirect.com/science/article/pii/S1364032115015506}
    \BIBentrySTDinterwordspacing

    \bibitem{nathCoherencyBasedSystem1985}
    R.~Nath, S.~S. Lamba, and K.~P. Rao, ``Coherency based system decomposition
    into study and external areas using weak coupling,'' \emph{IEEE Transactions
        on Power Apparatus and Systems}, vol. PAS-104, no.~6, pp. 1443--1449, Jun.
    1985.

    \bibitem{chowTimeScaleModelingDynamic1982}
    J.~H. Chow, Ed., \emph{Time-Scale Modeling of Dynamic Networks with
        Applications to Power Systems}, ser. Lecture {{Notes}} in {{Control}} and
        {{Information Sciences}}.\hskip 1em plus 0.5em minus 0.4em\relax
    {Berlin/Heidelberg}: {Springer-Verlag}, 1982, vol.~46.

    \bibitem{podmoreIdentificationCoherentGenerators1978}
    R.~Podmore, ``Identification of coherent generators for dynamic equivalents,''
    \emph{IEEE Transactions on Power Apparatus and Systems}, vol. PAS-97, no.~4,
    pp. 1344--1354, Jul. 1978.

    \bibitem{jonssonSystemProtectionScheme2004}
    M.~Jonsson, J.~Daalder, and M.~Begovic, ``A system protection scheme concept to
    counter interarea oscillations,'' \emph{IEEE Transactions on Power Delivery},
    vol.~19, no.~4, pp. 1602--1611, Oct. 2004.

    \bibitem{dateAggregationPropertiesLinearized1991}
    R.~Date and J.~Chow, ``Aggregation properties of linearized two-time-scale
    power networks,'' \emph{IEEE Transactions on Circuits and Systems}, vol.~38,
    no.~7, pp. 720--730, Jul. 1991.

    \bibitem{chowSingularPerturbationAnalysis1978}
    J.~H. Chow, J.~J. Allemong, and P.~V. Kokotovic, ``Singular perturbation
    analysis of systems with sustained high frequency oscillations,''
    \emph{Automatica}, vol.~14, no.~3, pp. 271--279, May 1978.

    \bibitem{tylooKeyPlayerProblem2019}
    M.~Tyloo, L.~Pagnier, and P.~Jacquod, ``The key player problem in complex
    oscillator networks and electric power grids: Resistance centralities
    identify local vulnerabilities,'' \emph{Science Advances}, vol.~5, no.~11, p.
    eaaw8359, Nov. 2019.

    \bibitem{pagnierInertiaLocationSlow2019}
    L.~Pagnier and P.~Jacquod, ``Inertia location and slow network modes determine
    disturbance propagation in large-scale power grids,'' \emph{PLOS ONE},
    vol.~14, no.~3, p. e0213550, Mar. 2019.

    \bibitem{machowskiPowerSystemDynamics2020}
    J.~Machowski, Z.~Lubosny, J.~W. Bialek, and J.~R. Bumby, \emph{Power System
        Dynamics: Stability and Control}, 3rd~ed.\hskip 1em plus 0.5em minus
    0.4em\relax {Hoboken, NJ, USA}: {John Wiley}, 2020.

    \bibitem{Cou23}
    R.~Courant, ``Ein allgemeiner satz zur theorie der eigenfunktionen
    selbstadjungierter differentialausdr\"ucke,'' \emph{Nachrichten von der
        Gesellschaft des Wissenschaften zu G\"ottingen}, vol.~1, pp. 80--84, 1923.

    \bibitem{Urs18}
    J.~Urschel, ``Nodal decompositions of graphs,'' \emph{Linear Algebra and its
        Applications}, vol. 539, pp. 60--71, 2018.

    \bibitem{Bam20}
    B.~Bamieh, ``A tutorial on matrix perturbation theory (using compact matrix
    notation),'' \emph{arXiv:2002.05001}, 2020.

    \bibitem{fritzschMatrixPerturbationTheory2021}
    J.~Fritzsch, M.~Tyloo, and P.~Jacquod, ``Matrix perturbation theory of
    inter-area oscillations,'' in \emph{2021 60th {IEEE} Conference on Decision
        and Control ({CDC})}.\hskip 1em plus 0.5em minus 0.4em\relax {IEEE}, dec
    2021.

    \bibitem{bergenStructurePreservingModel1981}
    A.~Bergen and D.~Hill, ``A structure preserving model for power system
    stability analysis,'' \emph{IEEE Transactions on Power Apparatus and
        Systems}, vol. PAS-100, no.~1, pp. 25--35, Jan. 1981.

    \bibitem{Welfonder_1989}
    E.~Welfonder, H.~Weber, and B.~Hall, ``Investigations of the frequency and
    voltage dependence of load part systems using a digital self-acting measuring
    and identification system,'' \emph{{IEEE} Transactions on Power Systems},
    vol.~4, no.~1, pp. 19--25, 1989.

    \bibitem{O_Sullivan_1996}
    J.~O{\textquotesingle}Sullivan and M.~O{\textquotesingle}Malley,
    ``Identification and validation of dynamic global load model parameters for
    use in power system frequency simulations,'' \emph{{IEEE} Transactions on
        Power Systems}, vol.~11, no.~2, pp. 851--857, may 1996.

    \bibitem{colettaLinearStabilityBraess2016a}
    T.~Coletta and P.~Jacquod, ``Linear stability and the braess paradox in
    coupled-oscillator networks and electric power grids,'' \emph{Physical Review
        E}, vol.~93, no.~3, p. 032222, Mar. 2016.

    \bibitem{pagnierOptimalPlacementInertia2019}
    L.~Pagnier and P.~Jacquod, ``Optimal placement of inertia and primary control:
    A matrix perturbation theory approach,'' \emph{IEEE Access}, vol.~7, pp.
    145\,889--145\,900, 2019.

    \bibitem{hornMatrixAnalysis2012}
    R.~A. Horn and C.~R. Johnson, \emph{Matrix Analysis}, 2nd~ed.\hskip 1em plus
    0.5em minus 0.4em\relax {Cambridge ; New York}: {Cambridge University Press},
    2012.

    \bibitem{sakuraiModernQuantumMechanics2017}
    J.~J. Sakurai and J.~Napolitano, \emph{Modern Quantum Mechanics}, 2nd~ed.\hskip
    1em plus 0.5em minus 0.4em\relax {Cambridge}: {Cambridge University Press},
    2017.

    \bibitem{maApplication2ndOrder2012}
    J.~Ma, T.~Wang, Z.~Wang, and J.~S. Thorp, ``Application of 2nd order matrix
    perturbation to compute power system inter-area oscillation modes considering
    uncertainties,'' in \emph{2012 {{IEEE Power}} and {{Energy Society General
                        Meeting}}}, Jul. 2012, pp. 1--6.

    \bibitem{yangMatrixperturbationtheorybasedOptimalStrategy2018}
    Y.~Yang, J.~Zhao, H.~Liu, Z.~Qin, J.~Deng, and J.~Qi, ``A
    matrix-perturbation-theory-based optimal strategy for small-signal stability
    analysis of large-scale power grid,'' \emph{Protection and Control of Modern
        Power Systems}, vol.~3, no.~1, p.~34, Nov. 2018.

    \bibitem{yiDiffusionConsensusWeakly2021}
    Y.~Yi, A.~Das, B.~Bamieh, Z.~Zhang, and S.~Patterson, ``Diffusion and consensus
    in a weakly coupled network of networks,'' \emph{IEEE Transactions on Control
        of Network Systems}, pp. 1--1, 2021.

    \bibitem{Haa18}
    F.~Haake, S.~Gnutzmann, and M.~Ku\'s, \emph{Quantum Signatures of Chaos}.\hskip
    1em plus 0.5em minus 0.4em\relax {Cham, Switzerland}: {Springer}, 2018.

    \bibitem{vonneumannUberVerhaltenEigenwerten1929}
    J.~{von Neumann} and E.~P. Wigner, ``\"uber das verhalten von eigenwerten bei
    adiabatischen prozessen,'' \emph{Physikalische Zeitschrift}, vol.~30, pp.
    467--470, 1929.

    \bibitem{tylooRobustnessSynchronyComplex2018}
    M.~Tyloo, T.~Coletta, and P.~Jacquod, ``Robustness of synchrony in complex
    networks and generalized kirchhoff indices,'' \emph{Physical Review Letters},
    vol. 120, no.~8, p. 084101, Feb. 2018.

    \bibitem{Min_2019}
    H.~Min and E.~Mallada, ``Dynamics concentration of large-scale
    tightly-connected networks,'' in \emph{2019 {IEEE} 58th Conference on
        Decision and Control ({CDC})}.\hskip 1em plus 0.5em minus 0.4em\relax {IEEE},
    dec 2019.

    \bibitem{chowPowerSystemCoherency2013}
    J.~H. Chow, Ed., \emph{Power System Coherency and Model Reduction}, ser. Power
    Electronics and Power Systems.\hskip 1em plus 0.5em minus 0.4em\relax {New
        York}: {Springer}, 2013, no. Volume 94.

    \bibitem{romeresNovelResultsSlow2013}
    D.~Romeres, F.~D{\"o}rfler, and F.~Bullo, ``Novel results on slow coherency in
    consensus and power networks,'' in \emph{2013 {{European Control Conference}}
        ({{ECC}})}, Jul. 2013, pp. 742--747.

    \bibitem{griggIEEEReliabilityTest1999}
    C.~Grigg, P.~Wong, P.~Albrecht, R.~Allan, M.~Bhavaraju, R.~Billinton, Q.~Chen,
    C.~Fong, S.~Haddad, S.~Kuruganty, W.~Li, R.~Mukerji, D.~Patton, N.~Rau,
    D.~Reppen, A.~Schneider, M.~Shahidehpour, and C.~Singh, ``The ieee
    reliability test system-1996. a report prepared by the reliability test
    system task force of the application of probability methods subcommittee,''
    \emph{IEEE Transactions on Power Systems}, vol.~14, no.~3, pp. 1010--1020,
    Aug. 1999.

\end{thebibliography}
\end{document}